# Quantum Readiness in Healthcare and Public Health: Building a Quantum Literate Workforce

Jonathan B. VanGeest, Ph.D.[1]; Kieran J. Fogarty, Ph.D.[2]; William G. Hervey, J.D., LL.M.[3]; Robert A. Hanson, D.O., M.S.[4]; Suresh Nair[5]; Timothy A. Akers, M.S., Ph.D.[6]

February 29, 2024

1. College of Public Health, Kent State University, Kent, Ohio.
2. College of Health Sciences, Western Michigan University, Kalamazoo, Michigan.
3. School of Health Sciences, Middle Georgia State University, Macon, Georgia; and the National Quantum Literacy Network, Baltimore, Maryland.
4. East Valley Cardiology, Chandler, Arizona; and the National Quantum Literacy Network, Baltimore, Maryland.
5. INA Solutions Inc., Fairfax, Virginia; and the National Quantum Literacy Network, Baltimore, Maryland
6. Office of Research Innovation and Advocacy, and School of Community Health and Policy, Public Health Program, Morgan State University, Baltimore, Maryland; and the National Quantum Literacy Network, Baltimore, Maryland.

Correspondence can be addressed to any of the individual authors:

- 1. J. VanGeest: jvangees@kent.edu
- 2. K. Fogarty: kieran.fogarty@wmich.edu
- 3. W. Hervey: bill.hervey@mga.edu
- 4. R. Hanson: hansonr44@gmail.com
- 5. S. Nair: suresh.nair@ina-solutions.com
- 6. T. Akers: timothy.akers@morgan.edu or takers@quantumliteracy.org

**Contributors**

All authors contributed equally to developing the concept and made substantial edits.

**Abstract**

*Quantum technologies, including quantum computing, cryptography, and sensing, among others, are set to revolutionize sectors ranging from materials science to drug discovery. Despite their significant potential, the implications for public health have been largely overlooked, highlighting a critical gap in recognition and preparation. This oversight necessitates immediate action, as public health remains largely unaware of quantum technologies as a tool for advancement. The application of quantum principles to epidemiology and health informatics—termed quantum health epidemiology and quantum health informatics—has the potential to radically transform disease surveillance, prediction, modeling, and analysis of health data. However, there is a notable lack of quantum expertise within the public health workforce and educational pipelines. This gap underscores the urgent need for the development of quantum literacy among public health practitioners, leaders, and students to leverage emerging opportunities while addressing risks and ethical considerations. Innovative teaching methods, such as interactive simulations, games, visual models, and other tailored platforms, offer viable*

*solutions for bridging knowledge gaps without the need for advanced physics or mathematics. However, the opportunity to adapt is fleeting as the quantum era in healthcare looms near. It is imperative that public health urgently focuses on updating its educational approaches, workforce strategies, data governance, and organizational culture to proactively meet the challenges of quantum disruption thereby becoming quantum ready.*

**The Dawning Quantum Age**

In 2023, the Cleveland Clinic unveiled the world's first quantum computer dedicated to healthcare research in Ohio, emphasizing the potential of quantum computing in revolutionizing the healthcare sector, particularly in drug discovery, genomics, and public health. By harnessing the principles of quantum theory, the IBM Quantum System One, installed at Cleveland Clinic's Lerner Research Center, an institute dedicated to developing a complete understanding of human health and disease, aims to reduce the critical timeline for new therapies and foster collaboration across pharmaceutical, biotech, genomics, and big data industries. This pioneering effort not only signifies a major step in technological advancement but also underscores Ohio's commitment to empowering the next generation of quantum literate health and public health talent and research in life sciences, with an expected economic impact of over $3 billion (Allchin, 2023, April 10).

Quite literally, both healthcare and public health are entering a new and evolving dawn. The principles of quantum mechanics may seem distant and abstract from a health perspective, but nothing can be further from the truth, as illustrated in the Clinic's recent purchase of the IBM Quantum System One. Quantum technologies encompass three subfields involving computing, communications, and sensing, with the ideas behind these subfields and their interaction holding significant potential for innovation and advancement. Quantum mechanics, which governs the behavior of subatomic particles and the interactions between light and matter, represents our deepest understanding of the micro-level workings of reality. The unconventional and counterintuitive probabilistic principles of quantum mechanics give rise to technologies that exceed by several orders of magnitude traditional, classical thinking and methods in accuracy, precision, and functionality.

Stepping back, we can see that quantum technologies are already upon us. These quantum technologies are already harnessing the health sciences through enhanced medical diagnostics, secure communication of sensitive health data, and rapid analysis of complex epidemiological information. By bridging the gap and fostering a connection between the quantum abstractions and the perceived grounding of both prevention and curative care, we may unlock new thinking and new avenues for improving health outcomes and well-being on a global scale.

For example, the binary states '0' and '1' can be likened to the states of non-infection and infection, respectively. Just as a classical bit in computing can only exist in one of two states, '0' or '1', an individual in a population can be simply classified as either non-infected ('0') or infected ('1') with a specific disease. However, when we move to the quantum realm, where qubits can exist in a superposition of both '0' and '1', the analogy becomes more complex and nuanced. This superposition can be likened to the varying degrees of infection risk across a population or subpopulation of interest. Some individuals might be at high risk of infection ('closer to 1'), while others may be at low risk ('closer to 0'), and many exist in a state of uncertainty between the two extremes.

This quantum analogy reflects the multifaceted, complex nature of infection risk, where factors like exposure, immunity, and health conditions create a continuum of risk rather than a simple binary, analog classification. Just as quantum computing leverages the superposition of states for more powerful calculations, a nuanced understanding of infection risk can lead to more targeted and effective interventions at both the individual and population levels. It is a rapidly evolving science at the intersection of individual and population health; uniquely equipping scientists to explore solutions unique to this transdisciplinary intersection. Quantum sensors, which exploit quantum entanglement, squeezing, and other exotic effects, could also provide unprecedented precision in measuring quantities like time, position, and electromagnetic fields, enhancing the accuracy of medical diagnostics and environmental monitoring (Degen et al., 2017).

In the field of biological sciences, quantum dots are highly effective because of their small size and the improved optical and electronic properties they exhibit due to quantum mechanics. They serve as passive quantum sensors to measure the concentration of bio-active molecules and are also employed as active quantum sensors in molecular and cellular imaging. Building on advances in quantum mechanics, recent cardiovascular research has developed innovative methods to precisely map the human heart's electrical conduction system, utilizing quantum sensors for enhanced accuracy. In the realm of cardiovascular research, decoding the neuro-cardiac axis is paramount and relies on the understanding of the electrical action potentials of the heart and brain domain. The advancing field of quantum cardiac research, more specifically the cardiac quantum spectrum, continues to promote cutting-edge technologies (Anderson et al. 2022).

As we step back from a health science perspective, especially from a practice level, we might distance ourselves because we perceive quantum physics as being confined to basic research at universities and national labs. However, quantum technologies are now transitioning rapidly from theoretical proposals into engineered commercial systems as their advantages and disruptive potentials become more apparent in the information age.

In recent years, key milestones like demonstrating quantum computational primacy over classical supercomputers using 53 functioning qubits have been achieved (Arute et al., 2019). Multiple vibrant startups have formed to compete in bringing quantum computing hardware, software, applications, and computational skills to the marketplace. Technology giants like IBM (and IBM-Q), Amazon, Microsoft, and Google are making major investments to offer cloud-based quantum computing access and resources. Total annual start-up investment is reaching all-time highs, year after year. Government science agencies and initiatives in the U.S., EU, China, and other regions are strategically funding quantum technology development and infrastructure, increasingly viewing quantum capabilities as vital for future economic competitiveness and national security.

This burgeoning "*second quantum revolution*" based on engineered quantum-enhanced devices will ultimately permeate and transform most technology sectors over the coming decades. However, quantum computing in particular is poised to confer immense potential benefits for data-driven and computationally intensive fields like healthcare and public health. Quantum simulation, optimization and machine learning techniques offer a path to accelerating analysis of massive, multipart datasets relevant to population health, allowing optimal surveillance, modeling, prediction and intervention across outbreaks, chronic diseases, environmental health, healthcare systems management and more (Dowling & Milburn, 2003). The data-centered evidence-based missions anchoring the practice of healthcare and public health positions both to be profoundly enhanced - but also disrupted - by the power of quantum computing, especially at the aforementioned but somewhat elusive effective melding of clinical- and community-based prevention and curative care.

**The Rise of Quantum Health Sciences**

Indeed, new cross-disciplinary domains synergistically combining quantum mechanical principles with health data, dubbed quantum health sciences, are rapidly emerging. Two prime examples are the related fields of *quantum health epidemiology* and *quantum health informatics*. Quantum health epidemiology utilizes the paradigm of quantum advantage, leveraging proprietary quantum algorithms and optimized hardware architectures to achieve exponential speedups in data processing and machine learning over classical computing systems. This allows elucidating disease and risk patterns, distributions, correlations, and causal mechanisms with greater resolution and timeliness to inform health policy and practice (Daley et al., 2022).

Quantum simulation can provide a virtual platform to rapidly model hypothetical infectious disease spread scenarios under varying conditions and assess the theoretical performance of countermeasures. Quantum machine learning techniques may likewise transform and accelerate knowledge discovery and forecasting across diverse population health and clinical datasets (Mesko, 2022). For example, by vastly improving our understanding of the implications of multiple – interrelated – factors such as urbanization, interaction patterns, vaccine-enabled and natural resistance, ecological changes, population health, health infrastructure, and more, we can better manage future pandemics (Dixon & Holmes, 2021).

Meanwhile, the domain of quantum health informatics is focused on prospectively applying anticipated exponential gains in data storage capacity, processing power, and machine learning from full-scale quantum computing to manage, process, model, analyze and extract insights from massive health and biomedical datasets (e.g., drug discovery, genomic analysis, medical imaging, etc.) for both individual and public benefit (Hackl & Ganslandt, 2017). Researchers in (*quantum*) epidemiology informatics envision that these advanced techniques will enhance clinical decision-making at the point of care, revolutionize the speed and success of pharmaceutical and biomedical research and development, optimize real-time responses to global pandemics, and enable ultra-personalized precision health interventions (Cossin & Thiébaut, 2020).

As shared, these innovations are made possible by harnessing orders of magnitude more health data for each individual, including information related to the genome, microbiome, and exposome. By improving the speed at which measurements are derived, as well as enhancing the probability of measuring correctly, we can create opportunities to maximize policy and practice response. This improvement also enhances our ability to predict future events based on pathogen modeling, and to assess the potential global response and costs associated with emergent and novel threats (Suciu, 2020).

While still emerging, active quantum information science research – with its focus on understanding how quantum physics principles can be utilized to process, store, and transmit information – has already begun demonstrating quantum proof-of-concepts on biomedical problems, including quantum computational modeling of COVID-19 outbreak dynamics assessing the impacts of various mitigation and control strategies (Padhi et al., 2020; Press, 2020). It also has the potential to help achieve a more sustainable healthcare system in the face of increased complexity and diversity of threats; further enhancing future public health response (Gupta et al., 2023). Recognizing this, technology leaders like IBM are now actively collaborating with major healthcare systems and national research institutes to prepare for a future when quantum health sciences alter the landscape of possible biomedical and public health research. Make no mistake, the time is swiftly approaching when quantum technologies will fundamentally transform healthcare. *The question remains, however, are the health sciences prepared for this shift?*

**The Urgent Need for Health and Public Health Quantum Literacy**

To date, the advanced physics, specialized mathematics, and abstract conceptual foundations underpinning quantum technologies have restricted proficiency largely to academic circles of elite theoretical physicists and engineers. Standard quantum curricula rely heavily on abstruse partial differential equations and linear algebra calculations that are disconnected from tangible intuition or applications for non-specialists. However, a range of creative interactive games, visual models, and virtual lab platforms (Foti et al., 2021) being developed show increasing promise for effectively demystifying the key principles of superposition, entanglement, and uncertainty underlying quantum systems in an intuitive way for non-physicists (Magnussen et al., 2019; Srivastava et al., 2022).

Making such interactive resources for cultivating functional quantum literacy widely accessible will be critical for health-related education and practice, as the novel integration of quantum principles and health data both enhances potential benefits but also exacerbates potential complications and risks requiring specialized oversight. For instance, storing and processing individuals' highly sensitive health records and biometrics on emerging quantum computing systems will necessitate upgraded data privacy and encryption safeguards to maintain public trust. The increasing complexity of quantum-enhanced epidemiological modeling and machine learning systems also requires comprehensive validation as well as transparency and interpretability to avoid misleading outputs or misuse. A workforce fluent in relevant quantum techniques will be essential for public health agencies and healthcare systems seeking to collaborate with the tech industry to co-design and implement quantum technologies ethically and equitably.

Therefore, strategically cultivating "quantum literacy" - meaning not just basic awareness but sufficient practical understanding and technological readiness among public health workforces to drive judicious adoption - must become an urgent priority across clinical and public health education, practice, policymaking, and leadership (Akers et al., 2022). All health professionals and students should attain adequate levels of quantum public health literacy aligned to their roles to operate effectively in a quantum-centric future. Leaders must become quantum literate to steward organizational adaptation. Researchers and analysts will leverage quantum tools daily. Communicators will translate quantum-driven insights to the public. Frontline clinicians may wield quantum-enhanced diagnostics and therapies. This extends far beyond hospitals and academic centers, to include the military, Department of Defense, Department of Health and Human Services, and the US Public Health Service, as well as other government and private agencies actively engaged in promoting public welfare. For the health sciences, the era of quantum disruption is no longer a distant speculation - it is an imminent reality requiring proactive preparation.

**Teaching Intuitive Quantum Health Literacy**

Overcoming existing barriers to engaging quantum literacy education will require utilizing customized interactive platforms spanning games, digital models, virtual labs and augmented reality tailored to health issues, contexts and systems of care, and diverse learning modalities. These tools can immerse learners in experiencing and exploring quintessentially quantum phenomena hands-on without formal prerequisites in advanced physics or mathematics.

For example, well-designed quantum games incentivize learning engagement via entertainment, curiosity, challenge, real world scenarios, and competition (Foti et al., 2021). Their rules and mechanics implicitly teach foundational quantum principles including superposition, entanglement, and measurement (Magnussen et al., 2019). Discovering winning strategies by interacting within simulated quantum worlds builds intuitive grasp. Practically speaking, health educators can leverage variables like scoring systems, social interactivity, and narrative context to make tacitly internalizing quantum behaviors more rewarding and enjoyable than rote memorization. Games lower the activation energy to start grappling with quantum concepts.

As the health sciences become more advanced through the emergence of quantum computing, existing technologies such as virtual and augmented reality environments will allow for recreating pivotal quantum experiments. Such experiments like the double slit diffraction or tests of Bell's Inequalities, in which even researchers can begin embracing not as physicists, but rather, as a form of health quantum literacy awareness. Interacting with on-screen photons, atoms, and measuring devices develops visceral experiential insights about uniquely quantum physical properties and phenomena difficult to convey verbally (Dehlinger & Mitchell, 2002). However, these simulated laboratories democratize exploring the quantum universe first-hand while, at the same time, have value in helping the health professions to become more quantum literate.

Moreover, interactive visual models represent abstract quantum information concepts like qubits, gates, and time evolution visually through animations rather than equations (Rieffel & Polak, 2022). Manipulating and observing such dynamic representations provides intuition about counterintuitive ideas like quantum parallelism and tunneling. While distilling the mathematical formalism, interactive conceptual models remain formally accurate. As flight simulators familiarize pilots experientially, quantum simulators foster functional intuitions about quantum behaviors or computational behaviors, such as in epidemiology, informatics, or statistics.

Blended learning, which combines interactive hands-on activities with expert narration, texts, and some lightweight equations/formulas, can frame progressive quantum literacy appropriate to health-specific contexts and needs. The aim is to cultivate the essential literacy required to catalyze more advanced study at the frontier of practice, epidemiology, and informatics. This approach is not meant to replace rigorous physics education to create quantum engineers; rather, evaluations suggest heightened learner engagement and retention versus traditional passive quantum instruction (Stilgoe et al., 2021). Overall, creativity in utilizing multimedia tools promises to foster the health workforce's necessary quantum health literacy.

**Crucial Near-Term Quantum Applications in the Healthcare and Public Health**

While interactive platforms help convey conceptual foundations (such as online tools, simulations, or visualizations), highlighting practical quantum applications that are germane to healthcare and public health makes their relevance more tangible for audiences and clarifies the urgency of proactive literacy initiatives. Current quantum computers remain limited but are steadily improving and already demonstrating advantages over classical systems for specialized problems and proofs-of-concept. Some of the most transformative near-term quantum applications include:

• **Quantum Machine Learning**: By processing massive training datasets exponentially faster, quantum machine learning techniques offer a powerful tool for finding patterns in data (Biamonte et al., 2017). It is also a path to revolutionizing pattern recognition and forecasting for healthcare and public health (Lloyd et al., 2021). This could enable breakthroughs like ultra-accurate infectious disease prediction, optimized healthcare logistics, precision medicine, and more. This includes interventions at the intersections of the health professions. Quantum machine learning can also expedite the process of analyzing protein unfoldment.

• **Quantum Sensing**: By harnessing quantum coherence and entanglement, quantum sensors measure quantities such as electromagnetic fields with unparalleled sensitivity and resolution. This technology has broad applications, ranging from medical imaging to detecting microbes (Degen et al., 2017). In healthcare and public health, these sensors could be crucial for early detection and diagnosis, significantly reducing the cost and enhancing the accessibility of cancer diagnosis.

• **Quantum Cryptography**: Unbreakable quantum cryptographic systems permit perfectly secure transmission of sensitive health data between organizations and individuals to safeguard privacy (Fawaz et al., 2021; Ladd et al., 2010). The transition to "post-quantum" cryptography was outlined as a priority for the Department of Homeland Security in 2021, as part of a roadmap to help organizations protect data and systems via quantum computing technology (DHS, 2021). This ensures the confidentiality and integrity of health information, a critical aspect of clinical and public health management.

• **Quantum Simulation**: Quantum computers can efficiently simulate molecular-level phenomena intractable for classical supercomputers to enable accelerated biomedical and biopharma advances (Langione et al., 2019; Popkin, 2019; Reiher et al., 2017). Such simulations can lead to the development of new drugs and treatments, directly impacting care and health outcomes.

Appreciating such concrete applications makes the value proposition of health quantum literacy more compelling and urgent for both professionals and students anticipating long careers where quantum technologies may be commonplace. Further exponential quantum algorithmic and hardware improvements expected over coming decades will continue expanding this technology's disruptive potential across healthcare and medicine.

**Table 1** provides a more detailed matrix of the relationship between classical and quantum computing with respect to quantum health epidemiology and quantum health informatics and their impact on emerging quantum applications.

**Table 1: Classical and Quantum Computing from a Public Health Perspective**

| | Classical Computing | Quantum Computing |
|---|---|---|
| **(Quantum) Health Epidemiology** | 1. **Simulation Data:** Traditional computational methods in epidemiology involve simulating disease models on classical computers. For complex models with a large number of parameters, this could be computationally expensive. Detailed population data, disease parameters, environmental factors used to model disease spread.<br>2. **Data Analysis:** Data analysis methods such as statistical regression, time series analysis, and machine learning are used to analyze and predict disease progression and trends.<br>3. **Epidemiological Data:** Disease prevalence and incidence data, demographic data, and other health-related data for statistical analysis and prediction.<br>4. **Disease Modelling:** Data on disease characteristics, demographics, and environmental factors are fed into simulations that model the spread of a disease, helping to predict and strategize responses.<br>5. **Disease Trend Analysis:** Incidence and prevalence data, as well as other health-related statistics, are analyzed to identify disease trends and to support clinical and public health decision-making. | 1. **Quantum Optimization Parameters:** Utilizing the application of quantum principles of superposition and entanglement, quantum computers could potentially solve optimization problems - such as determining the optimal intervention strategies in disease outbreaks - more efficiently than classical computers.<br>2. **Quantum Machine Learning:** Quantum algorithms could potentially improve the *speed and accuracy* of machine learning tasks in epidemiology, such as predictive modeling and anomaly detection.<br>3. **Predictive Data:** Quantum machine learning algorithms could potentially *generate new types of predictive data*, such as probabilities that consider quantum superposition of a disease super-spread.<br>4. **Enhanced Modelling:** Quantum computers may offer *faster and more complex modeling of diseases*, potentially improving prediction accuracy and treatment/response strategy.<br>5. **Advanced Trend Analysis:** Quantum machine learning could offer *improved analysis of disease trends*, potentially identifying patterns and making predictions that were not feasible with classical approaches. |
| **(Quantum) Health Informatics** | 1. **Information Processing:** Classical information theory forms the basis of our current understanding of communication, coding, and data compression in the digital world.<br>2. **Algorithms:** Classical algorithms serve as the foundation of many modern software applications and computational tasks. They are also used to solve problems in informatics, such as searching, sorting, and optimization.<br>3. **Information Data:** Digital information, such as text, images, and videos, that is encoded, transmitted, and decoded using classical information theory.<br>4. **Algorithm Inputs and Outputs:** Depending on the problem being solved, classical algorithms can work with a wide variety of data types, from simple numbers and strings to complex data structures.<br>5. **Information Management:** Digital information like patient records, images, and clinical data are managed, encoded, transmitted, and decoded in a secure and efficient manner using classical computing techniques.<br>6. **Data Processing:** Traditional algorithms are used for searching, sorting, and managing health data, as well as for problem-solving within healthcare informatics. | 1. **Quantum Cryptography:** Quantum principles like no-cloning (e.g., it is impossible to create an identical copy of an arbitrary unknown quantum state) and quantum entanglement enable the creation of un-hackable communication channels, enhancing the security of health information transfer.<br>2. **Quantum Algorithms:** Quantum computers have the potential to solve certain problems much more efficiently than classical computers. Examples include Shor's algorithm for factorization and Grover's algorithm for database search.<br>3. **Quantum Keys:** In quantum cryptography, the keys used for encryption and decryption are typically random sequences of qubits.<br>4. **Quantum States:** The input to a quantum algorithm is typically a quantum state, which can be a superposition of many different states. The output can also be a quantum state, which must then be measured to obtain a classical result.<br>5. **Improved Security:** Quantum cryptography could provide heightened security for sensitive health information, making the transfer and storage of such data less susceptible to breaches.<br>6. **Enhanced Processing:** Quantum algorithms may perform data processing tasks faster and more efficiently, potentially enabling new insights and capabilities in healthcare informatics. |

**Building a Quantum-Ready Health Workforce**

Adequately preparing the health workforce to assimilate the forthcoming quantum-driven transformations in healthcare will require updating curriculum, credentials, and career development programs across student and professional settings. Industry demand for quantum-related skills already exceeds supply in many developed countries, and global quantum specialized labor shortages may intensify through the 2020s absent proactive education policies (Kaur et al., 2022; Masiowski, et al., 2022).

For students, new undergraduate and graduate degree programs in quantum information science and engineering have proliferated over the last decade, often combining physics, computer science, mathematics, and electrical engineering (Aiello et al., 2021; Office of Science, U.S. Department of Energy, 2021). However, specialized physics is insufficient alone for healthcare and public health. Holistic quantum literacy must involve both deep quantum experts and professionals with enough cross-training to collaborate interchangeably. Less intensive upskilling options through supplemental seminars, modules, and micro-credentials embedded within public health, biomedical and healthcare education programs will be equally critical for "*quantum readiness*" within the existing and emerging workforce (World Economic Forum, 2023; Akers et al., 2022; National Quantum Literacy Network, 2021). Both future graduates and current practitioners require tailored opportunities to attain sufficient baseline quantum health literacy without prohibitive advanced physics prerequisites.

Particularly crucial are proactive initiatives focused on significantly improving diversity by empowering women and minorities with equitable access to quantum education and career opportunities, given these groups' historical underrepresentation in physics and computing (Gibney, 2019). Their perspectives will be essential for stewarding quantum technologies' responsible and ethical development. Partnerships with government agencies, industry, philanthropy and community organizations can provide pipelines nurturing diverse quantum excellence across the health workforce (Phillips et al., 2022). Quite frankly, America's global quantum leadership depends on engaging all its human talent.

Overall, the banner of "*quantum literacy*" provides a strategic focal point for modernizing health education in alignment with this imminent technological shift. Pioneering institutions have introduced divisions like Stanford's Quantum Health Technologies to bridge this knowledge gap at the intersection of quantum information science and clinical care (QHTea, 2022). To avoid disruption, the health sciences must take similar steps now to future-proof its workforce.

**Toward a Quantum-Literate Health and Public Health Society**

Beyond specialized workforce upskilling, enhancing baseline quantum literacy more broadly across civil society will carry profound importance as quantum technologies permeate everyday life. Populations conversant in quantum principles are essential for wise public governance and policymaking as such rapid innovations transform society and raise novel legal, ethical, economic and social implications (Dowling & Milburn, 2003).

In addition, interactive games, activities, and media making quantum concepts accessible to the public can foster informed societal perspectives, dialogue and decision-making. Dynamic outreach programs in schools and communities combining interactive resources, enriching entertainment content, diverse role models and participatory problem-solving events centered on quantum themes will help catalyze and sustain interest across diverse demographics (Srivastava et al., 2022). Quantum literacy is a potent inoculation against polarization - it empowers collective benefits rather than dire futures. It also prepares individuals to enter into schools of medicine and public health more prepared to fully leverage technology to take on the complexities of the health issues faced today.

For practitioners on the frontlines, basic quantum fluency similarly promises to be indispensable knowledge as the fields' capabilities change. Whether enhancing surveillance, forecasting, informatics, interventions, or more, familiarity with quantum techniques can help maximize their augmentation while navigating risks. Ideally, policy leaders will be tasked with overseeing organizational adaptation and mission alignment with quantum-powered capabilities. Clinicians may soon wield quantum-enhanced diagnostics and therapeutics. Across roles, broad health quantum literacy is foundational to optimizing quantum applications for patients and populations while stewarding risks prudently.

**The Window to Adapt is Now**

Powerful quantum technologies will profoundly disrupt and reshape healthcare and public health over the coming years and decades in ways both foreseeable and yet unimaginable, as illustrated in Table 1. The peak capabilities may remain distant, but the slope is exponential. Preemptively transitioning education, workforce strategy, data governance, ethics policies and organizational culture is thus an urgent policy and practice imperative. Delaying action cedes initiative to early quantum adopters in other sectors or nations. Prioritizing health quantum literacy awareness now, through launching workforce development programs and co-making interactive learning platforms tailored to diverse practitioners and students, can equip coming generations to proactively channel quantum progress toward equitable collective benefit rather than passive disruption. With wisdom, vigilance and moral imagination as guides, an enlightened quantum health era beckons within humanity's grasp. But the clock is ticking, the curve is exponential, and the race is on to adapt. The health sciences must leave the starting gate.

**Limitations**

While this article provides an overview of the potential for quantum technologies to transform healthcare and public health, there are some limitations to note. First, many of the proposed quantum health applications are still in early theoretical or experimental stages. Real-world implementation at scale faces engineering challenges and will require continued hardware and software advances. Second, assessing the feasibility and tangible benefits of emerging quantum techniques relies heavily on mathematical models and simulations which require further validation. Third, we do not yet have robust evidence quantifying the improvements quantum approaches could provide over classical methods for specific use cases. Rigorously demonstrating quantum advantage will be an ongoing research priority. Fourth, this article focuses mainly on the technological possibilities without deeply examining the ethical, legal, accessibility and workforce implications of deploying quantum health innovations. Further scholarship should explore these human dimensions in greater depth. Finally, transforming healthcare and public health for the quantum era will incur significant costs for technology investment, infrastructure upgrades and workforce training which must be weighed relative to other pressing needs.

## Health Implications

If achieved, the quantum health innovations described could significantly enhance healthcare and public health practices including surveillance, forecasting, research, interventions and data privacy. Quantum-based diagnostics, therapies and digitized health management systems promise more personalized, predictive and preventative healthcare to improve population outcomes. However, thoughtfully assessing and addressing the risks as well as benefits will be critical. Public health agencies and health systems should begin strategically evaluating how emerging quantum technologies may impact their mission, developing "quantum readiness" plans (micro-credential certifications) for their workforces, data systems and ethical governance. Quantum literacy initiatives tailored for diverse practitioners and students can help demystify quantum principles and foster wise adoption. But quantum progress should not detract from addressing urgent current challenges. Policymakers must ensure quantum healthcare advances ultimately reduce, rather than exacerbate, disparities. With wise governance, quantum techniques offer tools to advance our nation's health goals but not a magic solution replacing human values and collective responsibility.

## Acknowledgments
The authors wish to thank members of the National Quantum Literacy Network (NQLN) for their review and input. To learn more about NQLN, go to www.quantumliteracy.org

## Conflicts of Interests
The authors have no conflicts of interest to declare.

## References

Akers, T., et al. (2022/2023). *National Quantum Literacy Network*. Quantum Literacy Magazines. Retrieved from www.quantumliteracy.org

Allchin, T. (2023, April 10). *Ohio Welcomes World's First Quantum Computer Dedicated to Healthcare Research*. Retrieved from https://www.jobsohio.com/blog/ohio-welcomes-worlds-first-quantum-computer-dedicated-to-healthcare-research.

Aiello, C. D. et al. (2021). Achieving a quantum smart workforce. *Quantum Science and Technology*, 6(3), 030501. https://doi.org/10.1088/2058-9565/abfa64

Andersson, M. P., Jones, M. N., Mikkelsen, K. V., You, F., & Mansouri, S. S. (2022). Quantum computing for chemical and biomolecular product design. *Current Opinion in Chemical Engineering*, *36*, Article 100754. https://doi.org/10.1016/j.coche.2021.100754

Arute, F., Arya, K., Babbush, R. *et al*. (2019). Quantum supremacy using a programmable superconducting processor. *Nature* **574**, 505–510. https://doi.org/10.1038/s41586-019-1666-5

Biamonte, J., Wittek, P., Pancotti, N. *et al*. (2017). Quantum machine learning. *Nature* **549**, 195–202. https://doi.org/10.1038/nature23474

Cossin, S., & Thiébaut, R. (2020). Public health and epidemiology informatics: Recent research trends moving toward public health data science. *Yearbook of Medical Informatics*, 29(1), 231-4.

Daley, A. J., Bloch, I., Kokail, C., Flannigan, S., Pearson, N., Troyer, M., & Zoller, P. (2022). Practical quantum advantage in quantum simulation. *Nature*, 607(7916), 667-676.

Degen, C. L., Reinhard, F., & Cappellaro, P. (2017). *Quantum sensing. Reviews of Modern Physics*, 89(3), 035002. https://doi.org/10.1103/RevModPhys.89.035002

Dehlinger, D., & Mitchell, M. W. (2002). Entangled photon apparatus for the undergraduate laboratory. *American journal of physics*, 70(9), 898-902.

Department of Homeland Security (2021). *Post-Quantum Cryptography*. Available at https://www.dhs.gov/quantum (Accessed August 1, 2023).

Dixon, B.E., & Holmes, J.H. (2021). Managing pandemics with health informatics. *IMIA Yearbook of Medical Informatics*, 30(1), 69-74.

Dowling, J. P., & Milburn, G. J. (2003). Quantum technology: the second quantum revolution. *Philosophical Transactions of the Royal Society of London. Series A: Mathematical, Physical and Engineering Sciences*, 361(1809), 1655-1674.

Fawaz, H., Shin, W., & Shin, S. Y. (2021). Secure quantum machine learning for healthcare data. *IEEE Journal of Biomedical and Health Informatics*, 25(8), 2521-2532.

Foti, C., Anttila, D., Maniscalco, S., & Chiofalo, M. L. (2021). Quantum Physics Literacy Aimed at K12 and the General Public. *Universe*, 7(4), 86. https://doi.org/10.3390/universe7040086

Gibney, E. (2019). What Einstein could not see. *Nature*, 566(7744), 285-287.

Gupta, S., Modgil, S., Bhatt, P.C., Jabbour, C.J.C., & Kamble, S. (2023). Quantum computing led innovation for achieving a more sustainable Covid-19 healthcare industry. *Technovation*, 120, 102544.

Hackl, W.O. & Ganslandt, T. (2017). Clinical Information Systems as the Backbone of a Complex Information Logistics Process: Findings from the Clinical Information Systems Perspective for 2016, *Yearb Med Inform* 2017; 26(01): 103-109 DOI: 10.15265/IY-2017-023 https://doi.org/10.15265/IY-2017-023

Kaur, M., & Venegas-Gomez, A. (2022). Defining the quantum workforce landscape: a review of global quantum education initiatives. *arXiv preprint arXiv*:2202.08940v3.

Ladd, T.D., Jelezko, F., Laflamme, R., Nakamura, Y., Monroe, C., & O'Brien, J.L. (2010). Quantum computing. *Nature*, 464, 45-53.

Lloyd, S. (2021). Quantum machine learning for data classification. *Physics*, 14, 79.

Langione, M., Bobier, J. F., Meier, C., Hasenfuss, S., & Schulze, U. (2019). *Will quantum computing transform biopharma R&D?* https://www.bcg.com/publications/2019/quantum-computing-transform-biopharma-research-development
12

Magnussen, S., Myers, E. B., & Glad, A. (2019, October). Teaching quantum computing through gamification and quantum games. In *Proceedings of the 20th Annual Conference on Information Technology Education* (pp. 91-96).

Masiowski, M., Mohr, N., Soller, H., & Zesko, M. (2022). *Quantum computing funding remains strong, but talent gap remains. McKinsey Digital.* Available at: https://www.mckinsey.com/capabilities/mckinsey-digital/our-insights/quantum-computing-funding-remains-strong-but-talent-gap-raises-concern (Accessed August 15, 2023).

Mesko, B. (2022). What can quantum computing do to healthcare? *The Medical Futurist*. https://medicalfuturist.com/quantum-computing-in-healthcare/

National Quantum Literacy Network. (2021). *Quantum Literacy*. Retrieved from https://www.quantumliteracy.org

Office of Science, U.S. Department of Energy. (2021). *Report of the Working Group on Educational Curriculum Needs for Quantum Information Sciences.* Retrieved from https://science.osti.gov/-/media/wdts/pdf/QIS-WG-Report.pdf

Padhi, A., Pradhan, S., Sahoo, P. P., Nayak, N. R., & Panigrahi, P. K. (2020). Studying the effect of lockdown using epidemiological modelling of COVID-19 and a quantum computational approach using the Ising spin interaction. *Scientific Reports*, 10(1), 1-12.

Popkin, G. (2019). Waiting for the quantum simulation revolution. *Physics*, 12, 112.

Press, G. (2020). Calling on AI and quantum computing to fight the coronavirus. *Forbes*, April 14. Accessed 8.10.2023

Rieffel, E. G., & Polak, W. H. (2022). *Quantum computing: A gentle introduction*. MIT Press.

Reiher, M., Wiebe, N., Svore, K.M., & Troyer, M. (2017). Elucidating reaction mechanisms on quantum computers. *PNAS*, 114(29), 7555-60.

Srivastava, A., Srivastava, V., & Chakraverty, S. (2022). Gamification and quantum physics: a review of educational games designed for teaching of quantum phenomena. *Physics Education*, 57(3), 035007. https://doi.org/10.1088/1361-6552/ac5ebb

Stilgoe, J., Lock, S. J., & Wilsdon, J. (2021). Why should we promote public engagement with science? *Public Understanding of Science*, 23(1), 4-15. doi:10.1177/0963662520958797

Suci, D. (2020). COVID-19, Epidemiology and Informatics. *Presentation at TTI/Vanguard's Transformed by Digital Conference*, Seattle, WA (https://www.youtube.com/watch?v=gTejXbNES4g). Accessed 4/19/2023.

World Economic Forum. (2023). *Quantum Readiness Toolkit*. Retrieved from https://www3.weforum.org/docs/WEF_Quantum_Readiness_Toolkit_2023.pdf